# Stellar Spectroscopy in the Near-infrared with a Laser Frequency Comb

Andrew J. Metcalf,[1,2] Tyler Anderson,[3] Chad F. Bender,[4,5] Scott Blakeslee,[4] Wesley Brand,[1,2] David R. Carlson,[1] William D. Cochran,[6] Scott A. Diddams,[1,2,*] Michael Endl,[6] Connor Fredrick,[1,2] Sam Halverson,[4,7] Dan D. Hickstein,[1] Fred Hearty,[4] Jeff Jennings,[1,2] Shubham Kanodia,[4,14] Kyle F. Kaplan,[5] Eric Levi,[4] Emily Lubar,[4] Suvrath Mahadevan,[4,14,**] Andrew Monson,[4] Joe P. Ninan,[4, 14] Colin Nitroy,[4] Steve Osterman,[8] Scott B. Papp,[1,2] Franklyn Quinlan,[1] Larry Ramsey,[4,14] Paul Robertson,[4,9] Arpita Roy,[4,10] Christian Schwab,[11] Steinn Sigurdsson,[4,14] Kartik Srinivasan,[12] Gudmundur Stefansson,[4,14] David A. Sterner,[4] Ryan Terrien,[1,13] Alex Wolszczan,[4,14] Jason T. Wright,[4] and Gabriel Ycas[1,2]

*1 Time and Frequency Division, National Institute of Standards and Technology, 325 Broadway, Boulder, CO 80305, USA*
*2 Department of Physics, University of Colorado, 2000 Colorado Avenue, Boulder, CO 80309, USA.*
*3 Department of Physics, Pennsylvania State University, University Park, PA 16802, USA*
*4 Department of Astronomy & Astrophysics, Pennsylvania State University, 525 Davey Lab, University Park, PA 16802, USA*
*5 Steward Observatory, University of Arizona, Tucson, AZ 85721, USA*
*6 Department of Astronomy and McDonald Observatory, University of Texas at Austin, Austin, TX 78712, USA*
*7 MIT Kavli Institute for Astrophysics, 70 Vassar St, Cambridge, MA 02109 USA*
*8 Johns Hopkins Applied Physics Lab, Laurel, MD 20723, USA*
*9 Department of Physics and Astronomy, University of California-Irvine, Irvine, CA 92697, USA*
*10 California Institute of Technology, 1200 E California Blvd, Pasadena, CA 91125, USA*
*11 Department of Physics and Astronomy, Macquarie University, Sydney NSW 2109, Australia.*
*12 Center for Nanoscale Science and Technology, National Institute of Standards and Technology, 100 Bureau Drive, Gaithersburg, MD 20899, USA*
*13 Department of Physics and Astronomy, Carleton College, Northfield, MN 55057, USA*
*14 Center for Exoplanets and Habitable Worlds, Pennsylvania State University, University Park, PA 16802, USA*

*\* scott.diddams@nist.gov*
*\*\* suvrath@astro.psu.edu*

**Abstract:** The discovery and characterization of exoplanets around nearby stars is driven by profound scientific questions about the uniqueness of Earth and our Solar System, and the conditions under which life could exist elsewhere in our Galaxy. Doppler spectroscopy, or the radial velocity (RV) technique, has been used extensively to identify hundreds of exoplanets, but with notable challenges in detecting terrestrial mass planets orbiting within habitable zones. We describe infrared RV spectroscopy at the 10 m Hobby-Eberly telescope that leverages a 30 GHz electro-optic laser frequency comb with nanophotonic supercontinuum to calibrate the Habitable Zone Planet Finder spectrograph. Demonstrated instrument precision <10 cm/s and stellar RVs approaching 1 m/s open the path to discovery and confirmation of habitable zone planets around M-dwarfs, the most ubiquitous type of stars in our Galaxy.

## 1. Introduction

Measurements of the periodic Doppler shift of a star using the spectroscopic radial velocity (RV) technique provide evidence of an unseen orbiting exoplanet and its minimum mass [1]. Complemented by photometric measurements of the transiting exoplanet [2], one can obtain the mass and radius (and density) of the exoplanet, which are critical parameters for classification and assessment of habitability. While the RV technique has been used extensively, the highest precision measurements have been limited to the visible region of the spectrum <700 nm, and not the infrared [3]. Thus, 70% of the stars in our galaxy—M-dwarfs, which primarily emit in the near infrared (NIR)—have largely been outside the efficient spectral grasp of the current generation of proven RV instruments [4]. Due to their proximity, abundance, and small radius and mass, M-dwarfs are very attractive targets in the search for habitable zone rocky planets with NIR spectroscopy. The habitable zone [5] of an M-dwarf is close-in to the star, such that an orbiting Earth-mass planet within this zone induces an RV shift on the order of 1 m/s [6]. This is ten times larger than the RV signature of Earth around the Sun, and significantly increases the detectability of such an Earth-mass planet. Stellar convection and magnetic activity (e.g. granulation and starspots) can obfuscate the small RV signature of terrestrial planets, but many of these noise sources are suppressed in the NIR compared to the optical [7]. High RV precision measurements of bright stars in the visible will benefit from complementary high RV precision NIR observations that better discriminate stellar activity from real planet signals [7,8].

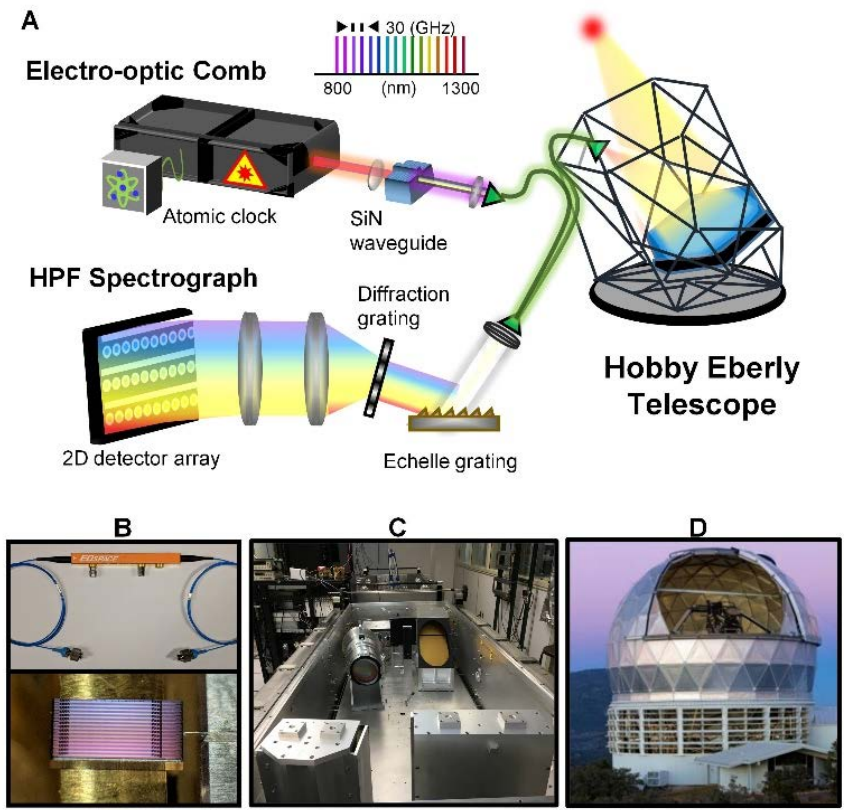

**Fig.1.** Instrumentation for precision infrared astronomical RV spectroscopy. (A) Starlight is collected by the Hobby-Eberly telescope and directed to an optical fiber. Lasers, electro-optics and nanophotonics are used to generate an optical frequency comb with teeth spaced by 30 GHz and stabilized to an atomic clock. Both the starlight and frequency comb light are coupled to the highly-stabilized Habitable Zone Planet Finder (HPF) spectrograph where minute wavelength changes in the stellar spectrum are tracked with the precise calibration grid provided by the laser frequency comb. (B) Components for frequency comb generation. (upper) A fiber-optic integrated electro-optic modulator and (lower) silicon nitride chip (5 mm × 3 mm) on which nanophotonic waveguides are patterned. Light is coupled into a waveguide from the left and supercontinuum is extracted from the right with a lensed fiber. (C) The HPF spectrograph, opened and showing the camera optics on the left, echelle grating on the right, and relay mirrors in front. The spectrograph footprint is approximately 1.5 m × 3 m. (D) The 10 m Hobby-Eberly telescope at the McDonald Observatory in southwest Texas.

While these motivations are well known, a combination of factors has limited the precision of even the best infrared RVs to 5-10 m/s over timescales of months [9,10]. Low-noise silicon detector arrays are not efficient at wavelengths beyond ~900 nm, requiring infrared detectors and cryogenic instruments. Existing wavelength calibration sources in the infrared are not yet as effective as iodine cells and thorium-argon lamps in the visible. Fiber-fed spectrographs with the required intrinsic stability in the infrared are only now being built and tested [11-13]. We overcome many technological and analysis hurdles that have impeded NIR RV measurements from reaching the ~1 m/s level, and we introduce a complete suite of infrared spectroscopic tools and techniques that provide the necessary precision for discovering and characterizing planetary systems around M-dwarf stars (see Fig. 1). We have developed a 30 GHz optical frequency comb spanning 700-1600 nm built on integrated electro-optic (EO) modulators [14-16] and high-efficiency nano-photonic nonlinear waveguides [17,18] to provide a robust calibrator for long-term operation at the telescope. This comb-based calibrator provides tailored light to the stable Habitable Zone Planet Finder (HPF) spectrograph, designed and built from the ground up for precision infrared RVs [19]. The frequency comb and spectrograph have been installed at the 10 m Hobby-Eberly Telescope, and we have demonstrated differential stellar RVs at 1.53 m/s (RMS) over months-long timescales, as well as shown that the comb-calibrated HPF can support RV precision as low as 6 cm/s.

To the best of our knowledge, this is the first time such RV precision has been realized in the near infrared wavelength region beyond the spectral grasp of silicon detector arrays. This interdisciplinary result is an important step on the path to discovery and characterization of Earth-mass planets in the habitable zones of the nearest stars. The higher planet-to-star radius ratio in these M dwarf systems will boost the transit signal by almost two orders of magnitude compared to Sun-like stars, making atmospheric studies of Earth-like planets feasible with current technology. Moreover, discovering and characterizing such targets is critical for the near-future detection of biomarkers in transiting terrestrial planetary atmospheres via spectroscopy with the James Webb Space Telescope [20].

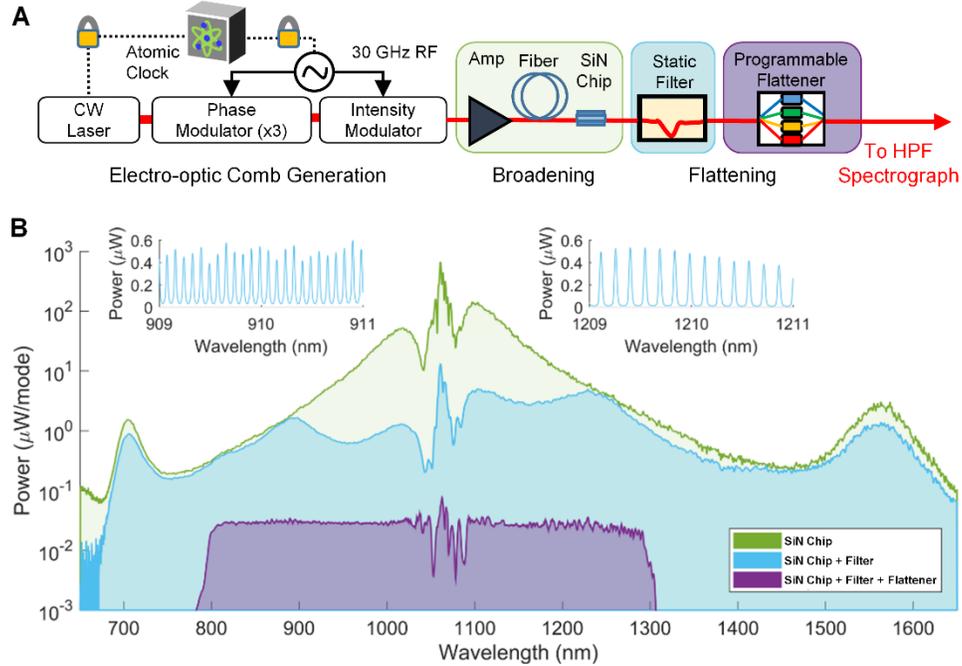

**Fig. 2.** The 30 GHz electro-optic frequency comb. (A) The frequency comb is generated via electro-optic modulation of a continuous wave (CW) laser followed by nonlinear spectral broadening and amplitude filtering stages to tailor the spectrum. (B) The spectral envelopes recorded with a low-resolution grating spectrometer at different points in the setup: Green - at the output of the SiN chip waveguide, Blue - after a static amplitude filter, and Purple - after a programmable amplitude filter. The two insets are high-resolution recordings of the 30 GHz comb modes centered at 910 and 1210 nm respectively.

## 2. Frequency Comb Instrumentation

Optical frequency combs [21], which consist of an equally spaced array of laser frequencies, have been recognized as a critical component in precision astronomical RV measurements aimed at identifying Earth-like exoplanets [9,22-26]. However, the combination of spectral coverage across hundreds of nanometers, line spacing of many tens of GHz (needed for modern high-resolution astronomical spectrometers), operational robustness in an astronomical observatory environment, and high up-time requirements are challenging to achieve. Multiple frequency comb approaches targeting astronomical spectrograph calibration have been explored, including mode filtering [9,22-26], electro-optic generation [15,27], and microresonators [28,29].

Our frequency comb is built around a combination of electro-optic and integrated- photonic technologies to address the challenges of bandwidth, mode spacing, and robustness; see Fig. 2A. In addition, we leverage recent advances with electro-optic frequency combs to mitigate the impact of multiplicative microwave phase noise on the comb tooth linewidth and coherence [16,18]. The frequency comb begins with 1064 nm continuous wave (CW) light from a semiconductor laser that feeds waveguide electro-optic modulators driven by a 30 GHz microwave source. This results in a comb of approximately 100 teeth spaced exactly by the microwave drive frequency. The CW laser, microwave source, and all other frequencies in the system are referenced to a GPS-disciplined clock that provides absolute traceability to the SI second. The initial comb is next passed through a 30 GHz Fabry-Perot cavity that acts to filter and reduce the impact of broad bandwidth electronic thermal noise [16]. The frequency comb is then amplified in an ytterbium fiber amplifier, spectrally broadened and temporally compressed to a pulse width of 70 fs, and finally focused into a 25 mm long nonlinear silicon nitride (SiN) waveguide [17,18]. The experimental setup is covered in more detail in Supplement 1.

The waveguide is 750 nm wide and 690 nm thick, which provides the combination of tight confinement and engineered dispersion to achieve a supercontinuum spectrum spanning 700-1600 nm with only 525 mW of incident average power (18 pJ of pulse energy). This low power reduces the thermal loading and aides the long-term operability. Following spectral generation, a combination of static and programmable amplitude filters [30] are used to tailor the spectral envelope. The output of each amplitude filtering stage is shown in Fig. 2B along with high resolution insets that show the resolved 30 GHz comb modes at 910 and 1210 nm. Optical heterodyne measurements between the 30 GHz comb and a frequency-stabilized reference laser confirm that the stability of the comb matches that of the GPS-disciplined clock with fractional uncertainty of $2\times10^{-11}$ at 1 s, and $<3\times10^{-13}$ for timescales of 1 day and longer. Significantly, electro-optic frequency combs could be employed for coverage from <600 nm to >2500 nm [17,18].

The entire frequency comb has been assembled on a 60 cm × 152 cm breadboard and installed at the Hobby-Eberly telescope at McDonald Observatory, together with the HPF spectrograph. The HPF is a vacuum-housed cross-dispersed echelle spectrograph with dimensions of approximately 1.5 m × 3 m and resolving power of $R=\lambda/\Delta\lambda\sim53,000$. The spectrograph is designed and optimized for stability, with its optics platform cryogenically cooled and temperature stabilized at the milliKelvin level [19,31]. The 28 echelle orders cover the wavelength range 810-1280 nm, and the spectra are recorded on a 2048×2048 pixel Hawaii-2 (H2RG) Mercury-Cadmium-Telluride (HgCdTe) infrared detector array that has a 1.7 μm long wavelength cutoff. The HPF entrance slit is fed by 3 optical fibers that can be simultaneously illuminated with star, sky, and calibration light to track the instrument drift. A more detailed description of experimental methods is provided in Supplement 1.

## 3. Spectrograph Calibration and Radial Velocity Measurements

The frequency comb was integrated into the HPF calibration system at the end of February 2018 and has been operating continuously and autonomously since early May 2018. The light from the comb system is coupled to the HPF calibration source selector system (see Kanodia *et al.* [32] for a layout) with a 50 μm diameter multimode fiber. This selector couples the light from any source (including the comb) into a 300 μm dynamic fiber agitator which is coupled to one port of a two-inch integrating sphere. We have shown [33] that the combination of temporal scrambling and an integrating sphere is effective at minimizing modal noise to deliver a constant illumination. The dynamic fiber-agitator is a customizable commercial system from Giga Concept that applies torsional oscillations to the fiber (under compression) at a few Hz. An identical second fiber agitator is also used to couple the calibration light to the calibration fiber coupled to the telescope facility calibration unit.

On-sky stellar observations of stable M-dwarfs and stars with known orbiting planets have been employed to characterize the system's present RV precision. Figure 3A shows example stellar and comb spectra as recorded on the HgCdTe array, illustrating the HPF's spectral range and the uniformity of the comb across the 28 orders. Using the comb as a reference, a wavelength solution is derived for each order that is subsequently used to calibrate the corresponding stellar spectra. Benchmarking the full instrumental chain (from telescope to comb to detector), as well as spectral extraction, RV measurement algorithms, and corrections of barycentric radial velocity requires on-sky observations of a star that is intrinsically stable, or has a known RV signal. Barnard's star is a bright nearby star of spectral type ~M4. While its RVs show a low-amplitude signal attributed to an exoplanet [35], it is still among the most stable M stars known. Figure 3B shows the residual RVs of Barnard's star from 118 high signal-to-noise measurements over a 3-month period. Observations were limited to 5-minute exposures to prevent saturation of the NIR detector. The scatter of individual (5-minute) RV measurements is 2.83 m/s, and when data within a ~1-hour HET observation window (or track) are binned to increase the signal to noise, the scatter is reduced to 1.53 m/s. This RV precision is unprecedented in the near infrared, and approaches that of the best measurements for this star with visible-band spectrometers (e.g. 1.23 m/s using the HARPS instrument [34]). A more detailed description of experimental methods is provided in Supplement 1.

Additional experiments allow us to probe the intrinsic calibration stability of the HPF spectrograph independent of stellar observations. To accomplish this, we take advantage of an auxiliary optical fiber that transmits the laser comb light to the primary focus of the telescope, where it is diffused and sent to the HPF spectrograph through both the "science" and "sky" fibers; Fig. 4A. In this arrangement, two sets of comb calibration spectra are recorded by HPF and processed to provide RV wavelength solutions from both fiber channels. As the stable frequency comb is common to the fibers, the retrieved RVs will reveal the slow drift of the HPF in each channel. However, the difference of the two channels should ideally be zero within the uncertainty of the photon noise. As such, this measurement allows us to carefully probe non-common-mode variations between the light paths from the two fibers to the detector array that would ultimately limit the RV precision. Such instabilities could arise from independent drifts in the optical fiber feeds, the spectrograph optics, systematics and noise in the HgCdTe detector array, or data reduction issues.

Measurements were taken over multiple nights and the absolute drift of the HPF is recorded in both channels every 5 or 10 minutes, with the results shown in Fig. 4B. As seen in these data, the drift of the HPF was about 5-7 m/s during a 5-10 hour measurement window, with larger daily excursions due to the liquid nitrogen fills. The difference of the RVs of the two channels is shown in 4C, revealing that the scatter of a single 10 minute differential drift measurement is at the 20 cm/s level. We further see that if we bin multiple measurements, the RV precision improves approximately as the square root of total measurement time to as low as 6 cm/s at 300 minutes, Fig. 4D. Furthermore, we observe no statistically significant drift in the difference of the RVs retrieved from the two fibers over 6 days.

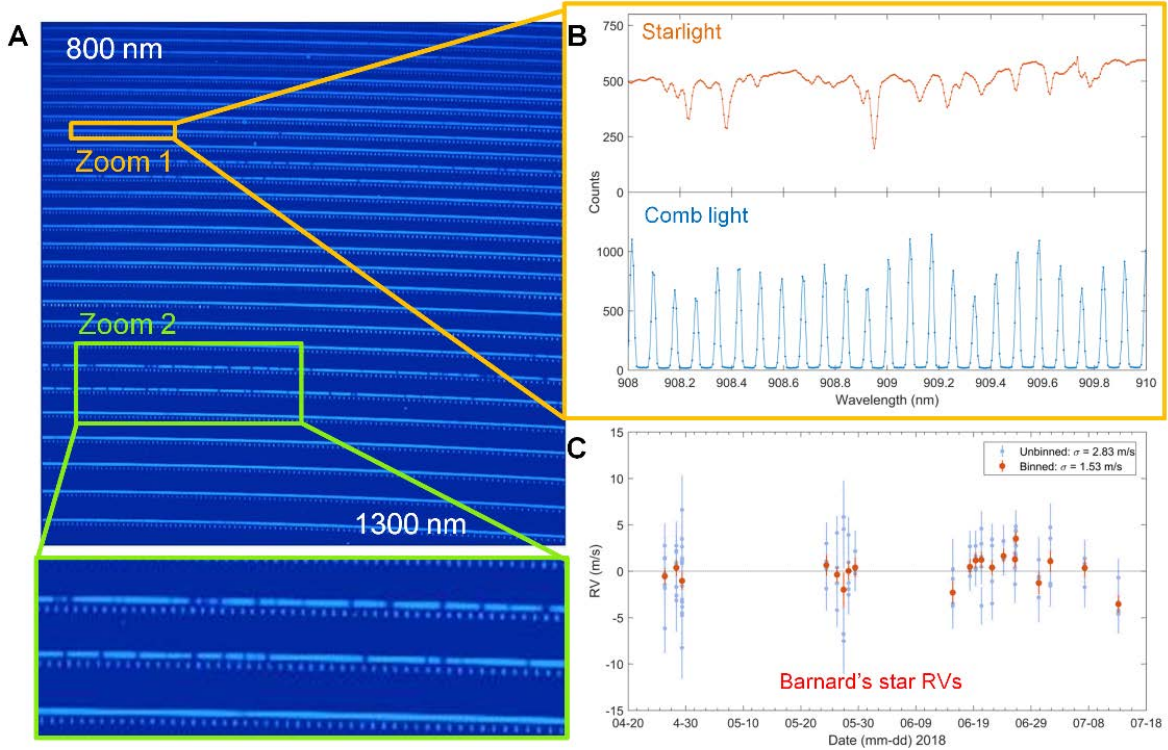

**Fig. 3.** On-sky data taken at the Hobby-Eberly telescope using the HPF spectrograph and laser frequency comb as a real-time calibrator. (A) Echellogram from the HPF detector array when illuminated by both the frequency comb and starlight from the telescope. The left image shows the full detector readout of the 28 echelle orders spanning 810-1280 nm. "Zoom 2" shows a smaller region where the vertical offset between the wavelength-matched star and comb light can be visualized. (B) "Zoom 1" shows the extracted stellar spectrum and comb calibration around 909 nm. (C) Three months of precision on-sky RV data of Barnard's star. Un-binned observations (5-minute cadence) are shown in blue. The binned observations are shown in red. The red points all have an equivalent on-sky exposure time of 20 minutes, or greater (see Table S1 in the Supplement for the binned RVs along with a listing of equivalent exposure times).

These data support the assertion that the comb-calibrated HPF has the intrinsic stability to support infrared spectroscopy with precision below 10 cm/s. With the HET this was a particularly harsh test since the calibration light splays across a significant fraction of the 30 arcmin focal plane—leading to fiber modal noise which is known to be worse in the NIR than the optical. A fiber-based mode scrambler was used to mitigate modal noise, but it likely does not fully solve this issue. Our measured RV scatter is currently about 5 times greater than the photon limited precision (3.7 cm/s in 10 minutes). Further improvements towards the photon limited precision should be possible with algorithmic improvements, better characterization of the NIR detectors, and better modal scrambling.

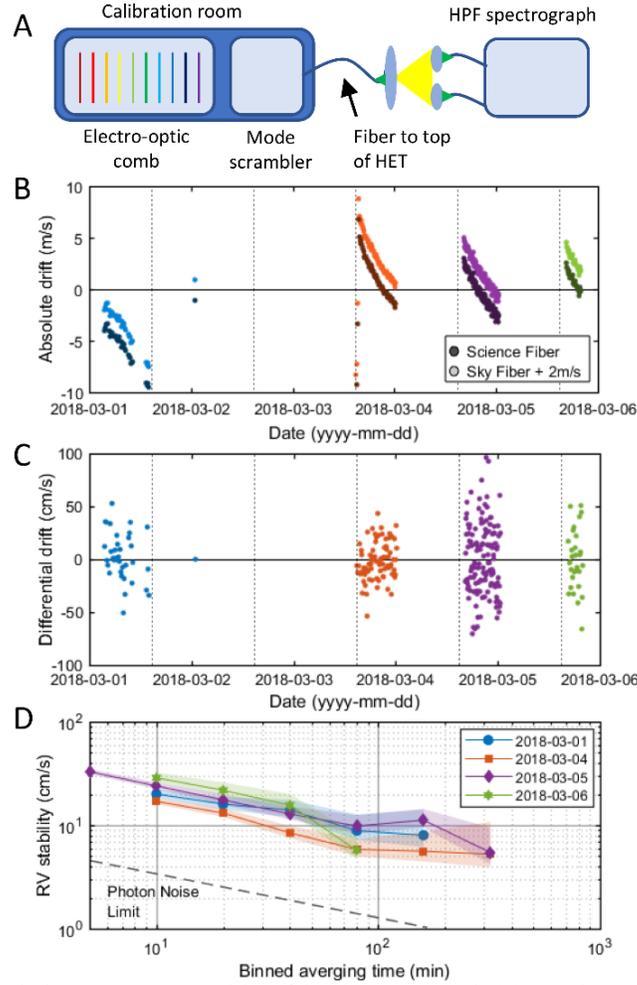

**Fig. 4.** Characterization of the radial velocity measurement precision of the HPF spectrograph. (A) Experimental setup. The laser frequency comb light is coupled via the HPF calibration bench and a fiber scrambler into an auxiliary optical fiber whose output is diffused across the HET focal plane to illuminate the HPF "science" and "sky" fibers. These fibers carry the light to the HPF spectrograph, where it is dispersed and detected on the HgCdTe detector array. (B) Absolute drifts of the HPF measured in the two fiber channels over 4 nights of tests. Each point is a 10 min exposure, except for data on 03-05, where the exposure time is 5 minutes. Note, the RVs from the "science" fiber are offset by 2 m/s in the plot for clarity. The total daily peak-to-peak drift of HPF is 15-20m/s, which is primarily caused by daily liquid nitrogen fills. Dashed lines show the time of these daily fills. The HPF was also exponentially approaching thermal stability in March when this test was performed. (C) The difference of the drifts between the RVs recovered from the "sky" and the "science" fibers. (D) Characterization of the RV stability by successive binning of measurements. This demonstrates that binning to effective averaging time of 300 minutes results in RV stability below 10 cm/s.

## 4. Discussion and Conclusion

In summary, we have introduced a new frequency comb platform in the NIR (700-1600 nm) that leverages the reliability of telecom electro-optic modulators and efficient chip-integrated nonlinear waveguides to provide an ultrastable calibrator for the HPF spectrograph. Since May 2018, the frequency comb and HPF instruments have been running autonomously at the 10 m Hobby-Eberly telescope, enabling high cadence RV measurements on one of the largest optical telescopes in the world. This combination of instrumentation forms a unique and powerful set of tools for exoplanet science focused on M-dwarfs. Our stellar RV measurements with scatter of 1.53 m/s are presently the most precise achieved in the NIR with HgCdTe detector arrays, and now approach the best RV measurements with more mature visible wavelength silicon detectors and associated technology. Within this context, the instrumentation and techniques we introduce represent a significant step towards the long-desired goal of RV spectroscopy in the NIR with precision at (and below) 1 m/s, as will be critical for the discovery and characterization of Earth-mass planets in the habitable zones of the nearest stars.

Finally, during the review of this paper, Ribas *et al.* [35] announced the discovery of a 3.2 Earth mass (RV amplitude = 1.2 m/s) exoplanet candidate orbiting Barnard's star with a period of 233 days. While our HPF RVs are not inconsistent with the orbit proposed by Ribas *et al.*, our observations extending over 86 days coincided with the least dynamic region of the orbital phase curve, resulting in the flat time series shown in Figure 3C. Nonetheless, it is significant that our first RVs from the comb-calibrated HPF are at a level of precision that makes them relevant for state-of-the-art exoplanet astrophysics. Further discussion of our RVs in the context of this exoplanet candidate is found in Supplement 1.

## Funding

National Science Foundation (NSF) (AST-100667, AST-1126413, AST-1310875, AST-1310885); NIST-on-a-Chip program, the Heising-Simons Foundation (grant #2017-0494), NASA NAI and NASA Origins grant (NNX09AB34G), Pennsylvania State University and its Center for Exoplanets and Habitable Worlds, NASA Earth and Space Science Fellowship Program (NNX16A028H). NASA Sagan Fellowship.

## Acknowledgments

We thank Zach Newman for his comments on this manuscript and Ian Coddington for his contributions to the development of the laser frequency comb. We thank the HET staff for their critical assistance, expertise and support. This work would not be possible without them. We are very grateful for help and support from Gary Hill, Hansin Lee, Brian Vattiat, and Phillip McQueen. The Center for Exoplanets and Habitable Worlds is supported by the Pennsylvania State University, the Eberly College of Science, and the Pennsylvania Space Grant Consortium. Data presented herein were obtained at the Hobby-Eberly Telescope (HET), a joint project of the University of Texas at Austin, the Pennsylvania State University, Ludwig-Maximilians-Universität München, and Georg-August Universität Gottingen. The HET is named in honor of its principal benefactors, William P. Hobby and Robert E. Eberly. The HET collaboration acknowledges the support and resources from the Texas Advanced Computing Center. Computations for this research were performed on the Pennsylvania State University's Institute for CyberScience Advanced CyberInfrastructure (ICS-ACI).

See Supplement 1 for supporting content.

## References

1. M. Mayor and D. Queloz, "A Jupiter-mass companion to a solar-type star," *Nature* **378**, 355 (1995)
2. Winn, J. N., "Exoplanet Transits and Occultations," in *Exoplanets*, S. Seager, Ed. (University of Arizona Press, Tucson, AZ, 2011), pp. 55-77.
3. D. A. Fischer, G. Anglada-Escude, P. Arriagada, R. V. Baluev, J. L. Bean, F. Bouchy, L. A. Buchhave, T. Carroll, A. Chakraborty, J. R. Crepp, R. I. Dawson, S. A. Diddams, X. Dumusque, J. D. Eastman, M. Endl, P. Figueira, E. B. Ford, D. Foreman-Mackey, P. Fournier, G. Fűrész, B. S. Gaudi, P. C. Gregory, F. Grundahl, A. P. Hatzes, G. Hébrard, E. Herrero, D. W. Hogg, A. W. Howard, J. A. Johnson, P. Jorden, C. A. Jurgenson, D. W. Latham, G. Laughlin, T. J. Loredo, C. Lovis, S. Mahadevan, T. M. McCracken, F. Pepe, M. Perez, D. F. Phillips, P. P. Plavchan, L. Prato, A. Quirrenbach, A. Reiners, P. Robertson, N. C. Santos, D. Sawyer, D. Segransan, A. Sozzetti, T. Steinmetz, A. Szentgyorgyi, S. Udry, J. A. Valenti, S.X. Wang, R. A. Wittenmyer, and J. T. Wright, "State of the Field: Extreme Precision Radial Velocities," *Publications of the Astronomical Society of the Pacific*, **128**, 066001 (2016).
4. T. J. Henry, W.-C. Jao, J. P. Subasavage, T. D. Beaulieu, P. A. Ianna, E. Costa, and R. A. Méndez, "The Solar Neighborhood. XVII. Parallax Results from the CTIOPI 0.9 m Program: 20 New Members of the RECONS 10 Parsec Sample," *Astronomical Journal* **132**, 2360 (2006)
5. R. K. Kopparapu, R. Ramirez, J. F. Kasting, V. Eymet, T. D. Robinson, S. Mahadevan, R. C. Terrien, S. Domagal-Goldman, V. Meadows, R. Deshpande, "Habitable Zones around Main-sequence Stars: New Estimates,", *Astrophysical Journal* **765**, 131 (2013)
6. G. Anglada-Escude, P.J. Amado, J. Barnes, Z. M. Berdiñas, R. P. Butler, G. A. L. Coleman, I. de la Cueva, S. Dreizler, M. Endl, B. Giesers, S. V. Jeffers, J. S. Jenkins, H. R. A. Jones, M. Kiraga, M. Kürster, M. J. López-González, C. J. Marvin, N. Morales, J. Morin, R. P. Nelson, J. L. Ortiz, A. Ofir, S.-J. Paardekooper, A. Reiners, E. Rodríguez, C. Rodríguez-López, L. F. Sarmiento, J. P. Strachan, Y. Tsapras, M. Tuomi & M. Zechmeister, "A terrestrial planet candidate in a temperate orbit around Proxima Centauri," *Nature* **536**, 437–440 (2016).
7. R. C. Marchwinski, S. Mahadevan, P. Robertson, L. Ramsey, J. Harder, "Toward Understanding Stellar Radial Velocity Jitter as a Function of Wavelength: The Sun as a Proxy," *Astrophysical Journal* **798**, 63-xx (2015).
8. P. Robertson, S. Mahadevan, M. Endl, and A. Roy, "Stellar activity masquerading as planets in the habitable zone of the M dwarf Gliese 581," *Science* **345**, 440 (2014)
9. G. G. Ycas, F. Quinlan, S.A. Diddams, S. Osterman, C. Bender, B. Botzer, L. Ramsey, R. Terrien, S. Mahadevan, S. Redman, "Demonstration of on-sky calibration of astronomical spectra using a 25 GHz near-IR laser frequency comb," Opt. Express 20, 6631 (2012).
10. J. Bean, A. Seifahrt, H. Hartman, H. Nilsson, G. Wiedemann, A. Reiners, S. Dreizler, T. Henry, " The CRIRES Search for Planets Around the Lowest-Mass Stars. I. High-Precision Near-Infrared Radial Velocities with an Ammonia Gas Cell", *The Astrophysical Journal*, 713, 410 (2009).
11. A. Quirrenbach, P. J. Amado, J. A. Caballero, R. Mundt, A. Reiners, I. Ribas, W. Seifert, M. Abril, J. Aceituno, F. J. Alonso-Floriano, H. Anwand-Heerwart, M. Azzaro, F. Bauer, D. Barrado, S. Becerril, V. J. S. Bejar, D. Benitez, Z. M. Berdinas, M. Brinkmöller, M. C. Cardenas, E.

Casal, A. Claret, J. Colomé, M. Cortes-Contreras, S. Czesla, M. Doellinger, S. Dreizler, C. Feiz,M. Fernandez, I. M. Ferro, B. Fuhrmeister, D. Galadi, I. Gallardo, M. C. Gálvez-Ortiz, A. Garcia-Piquer, R. Garrido, L. Gesa, V. Gómez Galera, J. I. González Hernández, R. Gonzalez Peinado, U. Grözinger, J. Guàrdia, E. W. Guenther, E. de Guindos, H.-J. Hagen, A. P. Hatzes, P. H. Hauschildt, J. Helmling, T. Henning, D. Hermann, R. Hernández Arabi, L. Hernández Castaño, F. Hernández Hernando, E. Herrero, A. Huber, K. F. Huber, P. Huke, S. V. Jeffers, E. de Juan, A. Kaminski, M. Kehr, M. Kim, R. Klein, J. Klüter, M. Kürster, M. Lafarga, L. M. Lara, A. Lamert, W. Laun, R. Launhardt, U. Lemke, R. Lenzen, M. Llamas, M. Lopez del Fresno, M. López-Puertas, J. López-Santiago, J. F. Lopez Salas, H. Magan Madinabeitia, U. Mall, H. Mandel, L. Mancini, J. A. Marin Molina,D. Maroto Fernández, E. L. Martín,S. Martín-Ruiz, C. Marvin, R. J. Mathar, E. Mirabet, D. Montes,J. C. Morales, R. Morales Muñoz, E. Nagel, V. Naranjo, G. Nowak, E. Palle, J. Panduro, V. M. Passegger, A. Pavlov, S. Pedraz, E. Perez, D. Pérez-Medialdea, M. Perger, M. Pluto, A. Ramón, R. Rebolo, P. Redondo, S. Reffert, S. Reinhart, P. Rhode, H.-W. Rix, F. Rodler, E. Rodríguez, C. Rodríguez López, R. R. Rohloff, A. Rosich, M A. Sanchez Carrasco, J. Sanz-Forcada, P. Sarkis, L. F. Sarmiento, S. Schäfer, J. Schiller, C. Schmidt, J. H. M. M. Schmitt, P. Schöfer, A. Schweitzer, D. Shulyak, E. Solano, O. Stahl, C. Storz, H. M. Tabernero, M. Tala, L. Tal-Or, R.-G. Ulbrich, G. Veredas, J. I. Vico Linares, F. Vilardell, K. Wagner,J. Winkler, M.-R. Zapatero Osorio, M. Zechmeister, M. Ammler-von Eiff, G. Anglada-Escudé, C. del Burgo, M. L. Garcia-Vargas, A. Klutsch, J.-L. Lizon, M. Lopez-Morales,A. Ofir, A. Pérez-Calpena, M. A. C. Perryman, E. Sánchez-Blanco, J. B. P. Strachan, J. Stürmer, J. C. Suárez,T. Trifonov, S. M. Tulloch,W. Xu, "CARMENES: an overview six months after first light," *Proc.SPIE* **9908**, 9908-14 (2016).
12. Étienne Artigau, Driss Kouach, Jean-François Donati, René Doyon, Xavier Delfosse, Sébastien Baratchart, Marielle Lacombe, Claire Moutou, Patrick Rabou, Laurent P. Parès, Yoan Micheau, Simon Thibault, Vladimir A. Reshetov, Bruno Dubois, Olivier Hernandez, Philippe Vallée, Shiang-Yu Wang, François Dolon, Francesco A. Pepe, François Bouchy, Nicolas Striebig, François Hénault, David Loop, Leslie Saddlemyer, Gregory Barrick, Tom Vermeulen, Michel Dupieux, Guillaume Hébrard, Isabelle Boisse, Eder Martioli, Silvia H. P. Alencar, José-Diaz do Nascimento, Pedro Figueira, "SPIRou: the near-infrared spectropolarimeter/high-precision velocimeter for the Canada-France-Hawaii telescope," *Proc.SPIE* **9147**, 9147-13 (2014).
13. Takayuki Kotani, Motohide Tamura, Hiroshi Suto, Jun Nishikawa, Bun'ei Sato, Wako Aoki, Tomonori Usuda, Takashi Kurokawa, Ken Kashiwagi, Shogo Nishiyama, Yuji Ikeda, Donald B. Hall, Klaus W. Hodapp, Jun Hashimoto, Jun-Ichi Morino, Yasushi Okuyama, Yosuke Tanaka, Shota Suzuki, Sadahiro Inoue, Jungmi Kwon, Takuya Suenaga, Dehyun Oh, Haruka Baba, Norio Narita, Eiichiro Kokubo, Yutaka Hayano, Hideyuki Izumiura, Eiji Kambe, Tomoyuki Kudo, Nobuhiko Kusakabe, Masahiro Ikoma, Yasunori Hori, Masashi Omiya, Hidenori Genda, Akihiko Fukui, Yuka Fujii, Olivier Guyon, Hiroki Harakawa, Masahiko Hayashi, Masahide Hidai, Teruyuki Hirano, Masayuki Kuzuhara, Masahiro Machida, Taro Matsuo, Tetsuya Nagata, Hirohi Onuki, Masahiro Ogihara, Hideki Takami, Naruhisa Takato, Yasuhiro H. Takahashi, Chihiro Tachinami, Hiroshi Terada, Hajime Kawahara, Tomoyasu Yamamuro, "Infrared Doppler instrument (IRD) for the Subaru telescope to search for Earth-like planets around nearby M-dwarfs," *Proc.SPIE* **9147**, 9147-12 (2014).
14. A. J. Metcalf, V. Torres-Company, D. E. Leaird and A. M. Weiner, "High-Power Broadly Tunable Electrooptic Frequency Comb Generator," *IEEE Journal of Selected Topics in Quantum Electronics* **19**, 231-236 (2013).
15. X. Yi, K. Vahala, J. Li, S. Diddams, G. Ycas, P. Plavchan, S. Leifer, J. Sandhu, G. Vasisht, P. Chen, P. Gao, J. Gagne, E. Furlan, M. Bottom, E. C. Martin, M. P. Fitzgerald, G. Doppmann, C. Beichman, "Demonstration of a near-IR line-referenced electro-optical laser frequency comb for precision radial velocity measurements in astronomy." *Nature Communications* **7**, 10436 (2016).
16. K. Beha, D. C. Cole, P. Del'Haye, A. Coillet, S. A. Diddams, and Scott B. Papp, "Electronic synthesis of light," *Optica* **4**, 406-411 (2017).
17. D. R. Carlson, D. D. Hickstein, A. Lind, J. B. Olson, R. W. Fox, R. C. Brown, A. D. Ludlow, Q. Li, D. Westly, H. Leopardi, T. M. Fortier, K. Srinivasan, S. A. Diddams, and S. B. Papp, "Photonic-chip supercontinuum with tailored spectra for precision frequency metrology" *Phys. Rev. Applied* **8**, 014027 (2017).
18. D. R. Carlson, D. D. Hickstein, W. Zhang, A. J. Metcalf, F. Quinlan, S A. Diddams, S. B. Papp, "Ultrafast electro-optic light with sub-cycle control" to appear in *Science* (*2018*).
19. S. Mahadevan, L. W. Ramsey, R. Terrien, S. Halverson, A. Roy, F. Hearty, E. Levi, G. K. Stefansson, P. Robertson, C. Bender, C. Schwab, M. Nelson, "The Habitable-zone Planet Finder: A status update on the development of a stabilized fiber-fed near-infrared spectrograph for the for the Hobby-Eberly telescope," *Proc.SPIE* **9147**, 9147–10 (2014).
20. S. Seager, W. Bains, and R. Hu, "Biosignature Gases in $H_2$-dominated Atmospheres on Rocky Exoplanets," *The Astrophysical Journal* **777**, 95 (2013)
21. S.A. Diddams, "The evolving optical frequency comb," *JOSA B* **27**, B51 (2010).
22. M. T. Murphy, T. Udem, R. Holzwarth, A. Sizmann, L. Pasquini, C. Araujo-Hauck, H. Dekker, S. D'Odorico, M. Fischer, T. W. Hänsch, and A. Manescau, "High-precision wavelength calibration of astronomical spectrographs with laser frequency combs," *Mon. Not. R. Astron. Soc.* **380**, 839–847 (2007).
23. D. Braje, M. Kirchner, S. Osterman, T. Fortier, S.A. Diddams, "Astronomical spectrograph calibration with broad-spectrum frequency combs," *Eur. Phys. J. D* **48**, 57 (2008).
24. C.-H. Li, A. J. Benedick, P. Fendel, A. G. Glenday, F. X. Kartner, D. F. Phillips, D. Sasselov, A. Szentgyorgyi, and R. L. Walsworth, "A laser frequency comb that enables radial velocity measurements with a precision of 1 cm s$^{-1}$," *Nature* **452**, 610–612 (2008).
25. T. Steinmetz, T.Wilken, C. Araujo-Hauck, R. Holzwarth, T.W. Hänsch, L. Pasquini, A. Manescau, S. D'Odorico, M. T. Murphy, T. Kentischer,W. Schmidt, and T. Udem, "Laser frequency combs for astronomical observations," *Science* **321**, 1335–1337 (2008).
26. T. Wilken, G. Lo Curto, R. A. Probst, T. Steinmetz, A. Manescau, L. Pasquini, J. I. Gonzalez Hernandez, R. Rebolo, T. W. Hänsch, T. Udem, R. Holzwarth, "A spectrograph for exoplanet observations calibrated at the centimetre-per-second level," *Nature* **485**, 611 (2012).
27. E. Obrzud, M. Rainer, A. Harutyunyan, B. Chazelas, M. Cecconi, A. Ghedina, E. Molinari, S. Kundermann, S. Lecomte, F. Pepe, F. Wildi, F. Bouchy, T. Herr, "Broadband near-infrared astronomical spectrometer calibration and on-sky validation with an electro-optic laser frequency comb," arXiv:1808.00860v2 (2018).
28. E. Obrzud, M. Rainer, A. Harutyunyan, M. H. Anderson, M. Geiselmann, B. Chazelas, S. Kundermann, S. Lecomte, M. Cecconi, A. Ghedina, E. Molinari, F. Pepe, F. Wildi, F. Bouchy, T. J. Kippenberg, T. Herr, "A Microphotonic Astrocomb," arXiv:1712.09526 (2018).
29. M.-G. Suh, X. Yi, Y-H. Lai, S. Leifer, I. S. Grudinin, G. Vasisht, E. C. Martin, M. P. Fitzgerald, G. Doppmann, J. Wang, D. Mawet, S. B. Papp, S. A. Diddams, C. Beichman, K. Vahala, "Searching for Exoplanets Using a Microresonator Astrocomb," arXiv:1801.05174 (2018).
30. A M. Weiner, "Ultrafast optical pulse shaping: A tutorial review," *Optics Communications* **294**, 3669-3692 (2011).

31. G. K. Stefansson, F. R. Hearty, P. M. Robertson, S. Mahadevan, T. B. Anderson, E I. Levi, C. F. Bender, M. J. Nelson, A. J. Monson, B. Blank, S. P. Halverson, C. Henderson, L. W. Ramsey, A. Roy, C. Schwab, R. C. Terrien, "A Versatile Technique to Enable Sub-milli-Kelvin Instrument Stability for Precise Radial Velocity Measurements: Tests with the Habitable-zone Planet Finder," *Astrophysical Journal* **833**, 175 (2016).
32. S. Kanodia, S. Mahadevan, L. Ramsey, G. Stefansson, A. Monson, F. Hearty, S. Blakeslee, E. Lubar, C. Bender, J. Ninan, D. Sterner, A. Roy, S. Halverson, P. Robertson, "Overview of the spectrometer optical fiber feed for the Habitable-zone Planet Finder", *Proc.SPIE* 10702, 10702-6 (2018).
33. S. Mahadevan, S. Halverson, L. Ramsey, N. Venditti, "Suppression of Fiber Modal Noise Induced Radial Velocity Errors for Bright Emission-line Calibration Sources", *The Astrophysical Journal,* 786, 18 (2014).
34. G. Anglada-Escudé, R. P. Butler, "The HARPS-TERRA Project. I. Description of the Algorithms, Performance, and New Measurements on a Few Remarkable Stars Observed by HARPS", Astrophysical Journal Supplements, 200, 15 (2012).
35. I. Ribas, M. Tuomi, A. Reiners, R. P. Butler, J. C. Morales, M. Perger, S. Dreizler, C. Rodríguez-López, J. I. González Hernández, A. Rosich, F. Feng, T. Trifonov, S. S. Vogt, J. A. Caballero, A. Hatzes, E. Herrero, S. V. Jeffers, M. Lafarga, F. Murgas, R. P. Nelson, E. Rodríguez, J. B. P. Strachan, L. Tal-Or, J. Teske, B. Toledo-Padrón, M. Zechmeister, A. Quirrenbach, P. J. Amado, M. Azzaro, V. J. S. Béjar, J. R. Barnes, Z. M. Berdiñas, J. Burt, G. Coleman, M. Cortés-Contreras, J. Crane, S. G. Engle, E. F. Guinan, C. A. Haswell, Th. Henning, B. Holden, J. Jenkins, H. R. A. Jones, A. Kaminski, M. Kiraga, M. Kürster, M. H. Lee, M. J. López-González, D. Montes, J. Morin, A. Ofir, E. Pallé, R. Rebolo, S. Reffert, A. Schweitzer, W. Seifert, S. A. Shectman, D. Staab, R. A. Street, A. Suárez Mascareño, Y. Tsapras, S. X. Wang, G. Anglada-Escudé, "A candidate super-Earth planet orbiting near the snow line of Barnard's star", *Nature*, 563, 365 (2018).

# SUPPLEMENT 1

## 1. LASER FREQUENCY COMB

The laser frequency comb is arranged on a 2.5' x 5' optical breadboard that is covered by a light-tight optical enclosure. The breadboard and enclosure are housed inside a temperature controlled 'calibration' room which resides in the basement of the Hobby-Eberly telescope (HET). This room also houses the HPF calibration bench and source selector. Support and diagnostic equipment, including the computer that controls and monitors the frequency comb operation, are located directly outside of the calibration room inside a standard 19" electronics rack. Custom designed software and an internet connection allow remote monitoring and control of the frequency comb from off-site locations. Similarly, the Habitable Zone Planet Finder (HPF) spectrograph is housed in a separate temperature-controlled room and has its own electronics rack and computer system for control. The computer systems of the frequency comb and HPF are fully integrated into the HET's internal network and can be accessed remotely by the resident astronomer during operation.

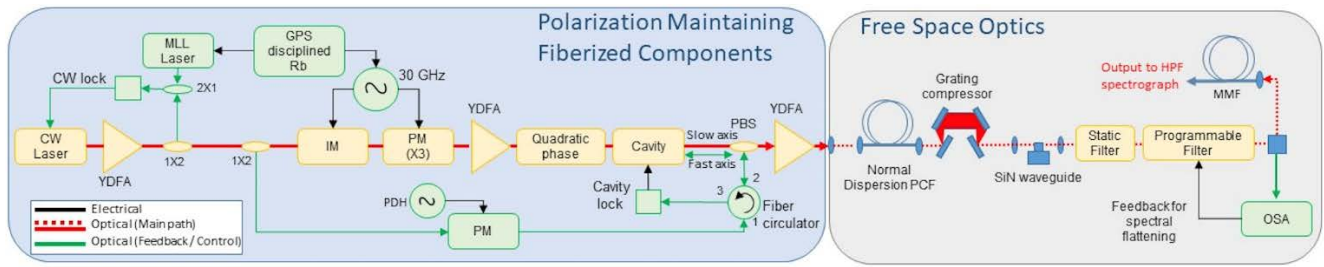

**Figure S1.** A detailed overview of the electro-optic frequency comb setup which is installed at the Hobby-Eberly telescope. (CW) continuous wave, (YDFA) ytterbium doped fiber amplifier, (IM) intensity modulator, (PM) phase modulator, (PDH) pound driver hall, (MLL) mode locked laser, (PBS) polarization beam splitter, (PCF) photonic crystal fiber, (SiN) silicon nitride, (OSA) optical spectrum analyzer, (MMF) multi-mode fiber out to HPF spectrograph calibration bench.

A detailed diagram of the frequency comb setup is shown in Fig. S1 and the basic architecture draws from recently-introduced techniques with electro-optics frequency combs [1-3]. A continuous wave (CW) laser at 1064 nm is amplified and then modulated at 30 GHz by three phase (PM) and one intensity modulator (IM) to generate ~100 comb lines spaced at 30 GHz. The number of PMs (three in this case) is chosen to provide the spectral bandwidth required to drive the subsequent nonlinear optical expansion of the spectrum. Only one IM is required, and it serves to carve out pulses with ~50% duty cycle from the phase-modulated light. When the phase of the 30 GHz signal applied to the IM and PM are correctly aligned in time, the IM will carve the light only when the chirp from the PMs is mostly linear, hence providing a relatively flat and easily compressed spectrum [1].

The CW laser is phase-locked to a home-built 125 MHz mode-locked laser (MLL) [4]. The MLL and 30 GHz RF signal are both phase-locked to a GPS-disciplined Rb clock to ensure the absolute frequency and spacing between the 30 GHz comb lines remain stable. The 30 GHz pulse train is then temporally compressed using pure quadratic phase from a programmable phase filter to achieve ~330 fs pulses. A linear Fabry-Perot (FP) optical cavity with finesse ~600 (linewidth of 50 MHz) is then used to reduce the multiplied thermal noise of the 30 GHz microwave source [2]. To align the comb and FP cavity for maximum transmission, a portion of the CW laser is split off from the main light path and modulated with PM sidebands to generate a Pound Drever Hall (PDH) error signal that is used to lock the cavity length to the CW light. The CW light is sent through the cavity in the reverse direction and on an orthogonal polarization state with respect to the main frequency comb light. The absolute frequency of the CW light relative to the 125 MHz reference comb is adjusted to maximize transmission of the 30 GHz comb. The 30 GHz comb output from the cavity is then amplified with ytterbium gain fiber to 2 Watts average power and output coupled to free space. Up to this point in the setup all components have been fiber-coupled and utilized polarization maintaining fiber.

The free space output from the amplifier is coupled into a 10 m length of normal dispersion photonic crystal fiber (PCF) to broaden the spectrum. Normal dispersion fiber was used to limit the effect of modulation instability and to provide a smooth spectral phase. At the output of the PCF the comb bandwidth spans ~50 nm and the smooth spectral phase allows subsequent recompression to a near band-limited duration of ~70 fs. A portion of the compressed pulse train (~525 mW) is focused into a silicon nitride (SiN) waveguide which produces a spectrum spanning from 700-1600 nm. The waveguide is silica cladded silicon nitride. The silicon nitride core has a cross section of 750 nm by 690 nm. These dimensions produce a region of anomalous dispersion around 1064 nm and appropriate phase matching to generate the dispersive waves near 700 nm and 1560 nm. The waveguides are tapered to 200 nm within 300 μm of the input and output to expand the mode profile for greater free space coupling efficiency. Ligentec fabricated the 25 mm long, low-loss waveguides. The resulting devices had up to 60% total throughput coupling using a diffraction limited input coupling lens.

We verify the stability and the linewidth of the 30 GHz comb following spectral broadening via heterodyne with independent stable CW lasers at 1070 nm, 1319 nm and 1565 nm. At these three wavelengths we observe comb tooth linewidths of 0.56 MHz, 0.955 MHz, and 1.4 MHz, respectively. The 0.56 MHz linewidth is consistent with the linewidth of the 1064 nm CW laser locked to the auxiliary 125 MHz comb, and the modest increase to 1.4 MHz at a frequency span of 91 THz (approximately 3000 comb teeth from the initial 1064 nm CW laser) is to be expected from microwave noise multiplication. However, this linewidth is much below the 50 MHz width of the Fabry-Perot filter cavity, indicating the efficacy of the high frequency cavity filtering on the ultimate comb linewidth. With the same optical heterodyne, we further verify that the center frequencies of the comb lines are precisely tied to the 30 GHz mode spacing and frequency of the 1064 nm laser with uncertainty better than $2 \times 10^{-11}$ for all times greater that 1 second. This is limited not by the frequency comb itself, but by the GPS-disciplined rubidium clock that serves as the master system frequency reference.

A static dielectric band reject filter is used to cut out the region of high optical power near the center of the broadened comb. The comb is then further tailored using a custom-built programable spectral amplitude filter. The filter utilizes a 2-dimensional liquid-crystal on silicon (LCoS) Spatial Light Modulator (SLM) and transmission grating arranged in a traditional 4-f pulse shaping setup [5]. The SLM has the ability to provide dynamic real-time control to update the filter mask to compensate for comb fluctuations. However, due to the relatively stable nature of the comb we have not yet utilized this feature and instead program the SLM with a static filter mask.

Light from the frequency comb is fed directly into the HPF calibration bench via a multimode optical fiber with a 62.5 μm diameter. From here, the comb light can be directed either to the HET Facility Calibration Unit [6] on the telescope through a Giga Concept fiber scrambler and auxiliary calibration fiber and used to illuminate the sky and science fibers of HPF or coupled through a second Giga Concept fiber scrambler which feeds an integrating sphere, with the HPF calibration fiber attached to one of the integrating sphere ports. These compact fiber scramblers help reduce modal noise from the highly coherent laser comb lines.

Our frequency comb architecture permits unsupervised operation with minimal remote intervention. Instead of nightly turn-on/turn-off procedures, the comb is kept on and ready to use at all times, and has an uptime of over 98% since continuous operation began more than 6 months ago. The majority of the downtime was due to a recurring computer issue that compromised the use of the comb for a total of 3 nights, but that has since been resolved. Building on the passive stability of the front-end components and the low-power supercontinuum, we have implemented automated computer control of servo set points for the comb, as well as control of the supercontinuum spectrum. Control software monitors the condition of each servo loop and automatically changes secondary parameters (e.g. temperature and powers) to keep the locks continuously in range of the primary servo output. We have also implemented slow feedback loops on the bias of the IM and the power injected into the SiN waveguide to maintain the spectral profile of the supercontinuum. With this, we have achieved long term relative amplitude fluctuations at the 10% level across the majority of the HPF bandwidth.

## 2. On-Sky Observations

The unique tilted Arecibo-style design of the HET allows observations of any given star only during one or two short tracks (~1hr) during the night. During such a track the constantly changing pupil of the HET leads to the variation in both collecting aperture and illumination. The use of octagonal fibers, with specially designed high-performance ball-lens scramblers [7] is necessary to achieve high RV precision. The LFC calibration light does not see this variable illumination. All observations at the HET are executed in a queue [8] by the HET resident astronomers.

As part of HPF early testing and commissioning to assess the on-sky RV stability of HPF, we obtained 191 spectra of Barnard's star (spectral type M4V, J band magnitude of 5.24) from February 25 to July 21 2018. Of these 191 spectra, we excluded a total of 25 spectra with a) a signal-to-noise ratio (SNR) <150 evaluated as the mean SNR at $\lambda$~1.07 micron (HPF order 57) since this was significantly lower than the median SNR (545) and indicated a problem with telescope guiding, acquisition or weather, b) spectra showing an excess sky-background and/or another stellar spectrum in the sky fiber, and c) spectra where a readout glitch introduced by a known issue with the Teledyne SAM H2RG GigE readout caused those spectra to be unusable.

This resulted in a total of 166 spectra with a median SNR of 545 per 1D extracted pixel at 1.07 micron. We used these 166 spectra for the generation of the master template, and performed a further cut of the spectra to present in the RV stream in Figure 3C, as is further described in the subsections below

### 2.1 Spectra used for the RV stream in Figure 3C

For the RVs presented in Figure 3C, we performed a further cut on the selection of the 166 spectra for the following reasons. First, the 166 spectra were taken in two overall different instrument and calibration system configurations which we describe here. Spectra taken from February 25-April 20 (45 spectra) were acquired as part of early HPF and LFC commissioning with varying instrument configurations (e.g., variations and upgrades to the calibration system fiber feed). Spectra taken from April 25-July 21 (121 spectra) were taken in the final configuration of both HPF and the calibration system. This time period is when we have the highest quality spectra, the most uniform drift calibration from the LFC, and regular flat field frames. We therefore only used spectra from the following time period for the RVs presented in Figure 3C.

Second, we only use tracks (the unique design of the HET only enables acquisition of targets during a discrete east or west track, ~1hr long for Barnard's star) that had 4 or more individual high-quality exposures per track to allow us to bin down RVs from observations in each track, with cumulative exposure time of at least 20 minutes (shown as the red points in Figure 3).

These additional cuts resulted in a selection of 118 spectra presented in the RV stream in Figure 3C. These spectra had a mean SNR of ~600 per 1D extracted pixel evaluated at $\lambda$~1.07 micron (HPF order 57). These spectra were obtained with two main exposure time lengths of either ~315s (99 spectra) or ~157.5s (19 spectra), corresponding to 30 and 15 up-the-ramp non-destructive reads on the HPF H2RG detector (10.45s for each read), respectively. The shorter exposure times were only performed on two nights (April 28-29), where the goal was to conduct higher cadence observations of Barnard's star while minimizing saturation. For all subsequent spectra, we adopted the longer exposure time for consistency, with the goal to achieve an SNR of 500-600 per spectrum per 1D extracted pixel at 1.07 micron.

### 2.2 Spectra used for the master RV template

To generate the highest SNR master template, we used the set of all 166 spectra from February 25–July 21, 2018. Although the early spectra from February 25–April 20 were not taken in the final configuration of the instrument, being at significantly different barycentric velocities than the later spectra, using them in the generation of the master template helped both generate a more accurate representation of the master template in the telluric regions, along with yielding an overall higher SNR template.

## 3. HPF Data Reduction

### 3.1 Generation of 1D extracted spectra

The full frame HPF H2RG array reads out non-destructively during exposures at a cadence of 10.45 seconds without any resets. Before fitting a slope to estimate e-/s flux from this up-the-ramp data, we corrected for various H2RG detector noise and artifacts. The largest source of correlated noise is bias fluctuations in the four HPF readout channels. Using the light-insensitive reference pixels and correlated signal in the light sensitive pixels between orders, we subtracted out the bias fluctuations to significantly below the readout noise. The bias-corrected data were then corrected for non-linearity using non-linearity curves we derived in laboratory testing. Detailed description of these algorithms can be found in Ninan et al. [9]. The 2D e-/s flux image calculated from the up-the-ramp data was then flat fielded to correct for pixel-to-pixel quantum efficiency differences. The echelle order position, as well as the profile on the array, was traced using a continuum source spectrum. This trace and profile were used for rectification and the optimal extraction of the flux to create a 1D spectrum [10-12]. For correcting the chromatic pixel sensitivity as well as other instrumental response across the spectrum, we divided the extracted spectrum using the de-blazed 1D extracted spectrum of the continuum source.

Wavelength calibration was done by fitting a smooth dispersion solution to the centroids of the LFC lines. The centroids were estimated by fitting individual Gaussian profile models with linear background. In typical LFC spectra, the background level is about 1% of the peak flux, and is primarily due to grating scattering. Our fitting algorithms, as described here, are largely robust against the background continuum at the present RV levels. For precise barycentric velocity as well as instrumental drift correction, flux weighted midpoints of the exposure time of each echelle orders were calculated using the differential flux in the raw up-the-ramp data.

For precise instrumental drift correction, we created a high signal-to-noise ratio template of the LFC from the epochs near the one used for solving the wavelength dispersion solution. This template was then transformed in pixel space by a polynomial model to fit the LFC spectrum at different epochs. The fit was performed by least square minimization of each echelle order individually. The residuals were weighted proportionally to the flux. However, to prevent the brightest LFC lines from dominating the minimization, we de-weighted the brightest 2 percentile pixels to the same level as the 98 percentile pixels. The instrumental drift measured from daily LFC comb calibration was then linearly interpolated to the flux-weighted epochs of stellar observations. This drift correction was then applied to the wavelength dispersion solution of each individual echelle order.

True instrumental RV precision is measured by calculating how well the instrumental drift can be corrected across the fibers in HPF. The measurement setup shown in Figure 4 illuminated both the science and the sky fiber of HPF simultaneously with the LFC. The instrumental drift was estimated separately from both the fibers as explained in the previous paragraph. The drift estimates were then subtracted from each other to measure any residual drift. Scatter in our estimate from a 10-minute exposure is 20 cm/s; this is larger than the expected photon noise level scatter of 3.7 cm/s. We have identified a few possible sources of errors like modal noise, and inter pixel sensitivity variation that need to be mitigated to further improve this measurement precision.

**3.2 RV reduction**

The RVs in Figure 3C were extracted with a least-square template-matching RV fitting method described by Anglada-Escude & Butler [13]. This algorithm is based on minimizing the differences of the observed spectrum against a high signal-to-noise master template constructed from a co-addition of all available observations of the target star. This method has been shown to be particularly powerful when extracting RVs for mid-to-late M-dwarf spectra, where the stellar continuum is not locally smooth due to the large number of molecular absorption lines [13]. To extract RVs from the HPF 1D spectra, we adapted the publicly available SpEctrum Radial Velocity Analyzer (SERVAL) code [14], a Python implementation of the RV template-matching algorithm. Here we provide an overview of how we extracted the HPF RVs, focusing on a brief description of the core algorithm along with highlighting our main changes to the base SERVAL code. We direct the interested reader to the SERVAL paper [14]. The template-matching algorithm in SERVAL calculates the RVs in three main steps. First, in the 'pre-RV' step, all of the spectra are first brought to the reference frame of the Solar system barycenter by correcting for the barycentric motion of the Earth. To calculate the barycentric velocity of the Earth we use the openly available barycorrpy package [15], which has been shown to be precise at the ~1cm/s level. After the barycentric correction, the highest SNR spectrum is used as an initial template to calculate the maximum likelihood 'pre-RV' through minimizing the differences between the observed spectra and the initial template. For the second step, the algorithm uses these 'pre-RVs' together with the barycentric correction to bring all of the individual spectra to a more accurate common reference frame to then coadd the individual spectra to create a high-SNR master template. As the third and final step, this master template is used to derive a more robust set of final RVs. For our RV reduction, we skipped the first 'pre-RV' step, i.e., we generated the template assuming a pre-RV of 0 for all of the spectra observed.

Although HPF has a total of 28 echelle orders covering the information-rich z, Y and J NIR bands, for our RV extraction we focused on using the 9 orders least affected by telluric absorption in the HPF bandpass. Specifically, for the extraction presented here we used spectra corresponding to grating orders 72-69 (covering the 841.15-889.5nm region in the z-band), and 61-57 (covering the 993.3-1076.7nm region in the Y-band). We are actively working on incorporating the other HPF orders for the RV extraction, to leverage the larger information content offered in doing so. However, more work remains on optimally extracting the RVs in the presence of large-scale telluric absorption regions.

To acquire the highest precision RVs, both lines due to telluric absorption and/or sky emission lines present in the atmosphere either need to be modeled out or explicitly masked. We follow the SERVAL implementation of both telluric and sky-emission lines masks: any areas affected by variable telluric and/or sky-emission are either severely down-weighted (in the creation on the master template spectrum), or fully masked out (in calculation of final RVs).

To generate a binary telluric mask, we first created a finely-sampled master telluric spectrum using the open source TelFit package [16]. TelFit is a Python wrapper to the well-known Line-By-Line Radiative Transfer Model package [17] which is capable of calculating a telluric-absorption spectrum given a number of different atmospheric input parameters. For our telluric mask, we calculated a synthetic telluric spectrum across the full HPF bandpass, assuming the following default TelFit values: a humidity of 50% at the HET latitude (30.6 degrees) and altitude (2100 m). We then used this output TelFit spectrum to generate a thresholded binary mask where we masked everything below a transmission of 99.5% as a telluric. We experimented with different thresholding levels from 99% to 99.9%, and use 99.5% for the best RV reduction, though in practice, the final standard deviation of the RVs did not vary substantially between these threshold levels. We elected to use the native high-resolution telluric spectrum from TelFit—i.e., without degrading the spectral resolution to the HPF R~53,000 mean resolution—to better help mask out smaller less-deep telluric regions. After creating the threshold binary mask, we further broadened the mask by 17 wavelength resolution elements (each resolution element in the mask being 0.027A). The exact choice of broadening did not impact the final RVs substantially.

**Table S1.** Binned radial velocity measurements and associated 1-sigma uncertainties from our observations of Barnard's star. These RVs are plotted in Figure 3C as the red points. The errors are the associated weighted-average errors from the 1-sigma un-binned errors from *SERVAL* (blue errorbars in Figure 3C), along with a fixed fiber-to-fiber LFC calibration error of 20cm/s per 5-minute bins added in quadrature. BJD and EET are Barycentric Julian Date and Equivalent Exposure Time, respectively.

| Time BJD | RV (m/s) | σRV (m/s) | EET (min) |
|---|---|---|---|
| 2458234.87746 | -0.44 | 0.82 | 41.4 |
| 2458236.87683 | 0.47 | 0.84 | 39.3 |
| 2458237.87357 | -0.94 | 0.91 | 36.7 |
| 2458262.94159 | 0.75 | 1.11 | 25.2 |
| 2458264.79527 | -0.27 | 0.82 | 25.3 |
| 2458265.93552 | -1.89 | 1.87 | 20.3 |
| 2458266.80063 | 0.13 | 0.95 | 25.2 |
| 2458267.94116 | 0.50 | 1.02 | 20.3 |
| 2458284.89181 | -2.21 | 1.18 | 25.2 |
| 2458287.87142 | 0.57 | 0.82 | 20.3 |
| 2458288.87240 | 1.27 | 0.84 | 20.3 |
| 2458289.86930 | 1.33 | 0.85 | 25.3 |
| 2458291.72307 | 0.50 | 0.79 | 25.3 |
| 2458293.71773 | 1.74 | 0.76 | 25.2 |
| 2458295.70465 | 1.37 | 0.84 | 25.2 |
| 2458295.84834 | 3.62 | 0.75 | 25.3 |
| 2458299.84102 | -1.17 | 1.15 | 25.3 |
| 2458301.84369 | 1.17 | 1.09 | 25.0 |
| 2458307.81320 | 0.46 | 0.94 | 25.2 |
| 2458313.66344 | -3.45 | 0.93 | 25.3 |
| 2458320.65061 | -1.44 | 0.95 | 25.2 |

We created a sky-emission line mask in a similar manner to the steps outlined above for the telluric masking with some slight differences. To create the mask we used a master sky-background frame taken by staring at a blank patch of the night sky for 10 minutes on the night of March 29th 2018 (moon phase of 98%). After subtracting the background continuum of this deep sky spectrum, we derived the locations of the stellar lines by thresholding anything above 5 sigma as a sky emission line. In addition to masking out the telluric and sky emission lines, we masked out stellar lines. We specifically masked out the 3 Ca II infrared triplet lines (Ca IRT) centered at 8498.02A, 8542.09A, and 8662.14A (air wavelengths), which are known to be highly sensitive to stellar activity [18]. In addition to these lines, we masked out the following 8 deep (>30% deeper than surrounding pseudo-continuum) atomic lines (8426.504A, 8434.959A, 8435.650A, 8468.4069A, 10496.122A, 10661.628A, 10677.053A, and 10726.392A; air wavelengths) that were either close to the edge of an order, and/or showed excess correlated noise (at the ~0.5-1% level) in the residuals. Although Barnard's star has been shown to exhibit low levels of stellar activity [19], we performed tests with ($\sigma$RV,binned=1.53m/s) and without these lines masked ($\sigma$RV,binned=1.77m/s). We additionally masked out three other regions—corresponding to 10140.2-10142.7A, 10527.7-10531.3A, and 10739.4-10742.2A (in the barycentric rest-frame)—that did not correspond to any deep lines but showed excess correlated noise in the residuals which we attribute as bad-pixel regions on the HPF detector. We continue to explore the reasons why these lines and bad-pixel regions add noise in an effort to disambiguate between detector and extraction induced issues and astrophysical noise. We surmise that improved flat fielding has the potential to further increase the overall RV precision presented here. These challenges, some from the unique design of the HET Facility Calibration Unit [8], remain to be overcome.

We have calculated the expected photon noise for the spectrum in the bandpass used using the formalism presented by [20]. We calculate the quality factor (Q) from the high signal-to-noise master RV template. For each observed spectrum we mask out tellurics and sky emission regions, along with the noisy stellar lines. Following [20], using the Q calculated from the template for the unmasked regions, we estimate the photon noise limited stellar RV precision (including detector read noise and digitization noise) per observation by taking a weighted average of the photon noise RV precision for each individual order. In doing so, we estimate a median photon noise RV precision of 1.68m/s compared to the as-observed scatter of 2.83m/s for the unbinned data. Similarly, for the binned data, we estimate a median photon noise error of 0.71m/s compared to the as-observed scatter of 1.53m/s. In both cases, we see that the as-observed scatter is larger by a factor of ~2, suggesting that our observations are limited by other noise sources.

Table S1 lists the weighted-average RVs (red points in Figure 3C), where each point represents the weighted-average radial velocity from all RVs from one HET track with a cumulative exposure time of at least 20 minutes. The errors are the associated weighted-average errors from the 1-sigma un-binned errors from SERVAL (blue errorbars in Figure 3C), along with a fixed fiber-to-fiber LFC calibration error of 20cm/s per 5-minute bins added in quadrature. The binning of RV observations obtained inside the same HET track (<1hr of each other) serves to increase the SNR and has historically also been used to beat down any short-term stellar activity and oscillations. We chose to acquire individual 5-minute observations and bin the RVs to avoid detector saturation or large non-linearity that would result from longer exposures.

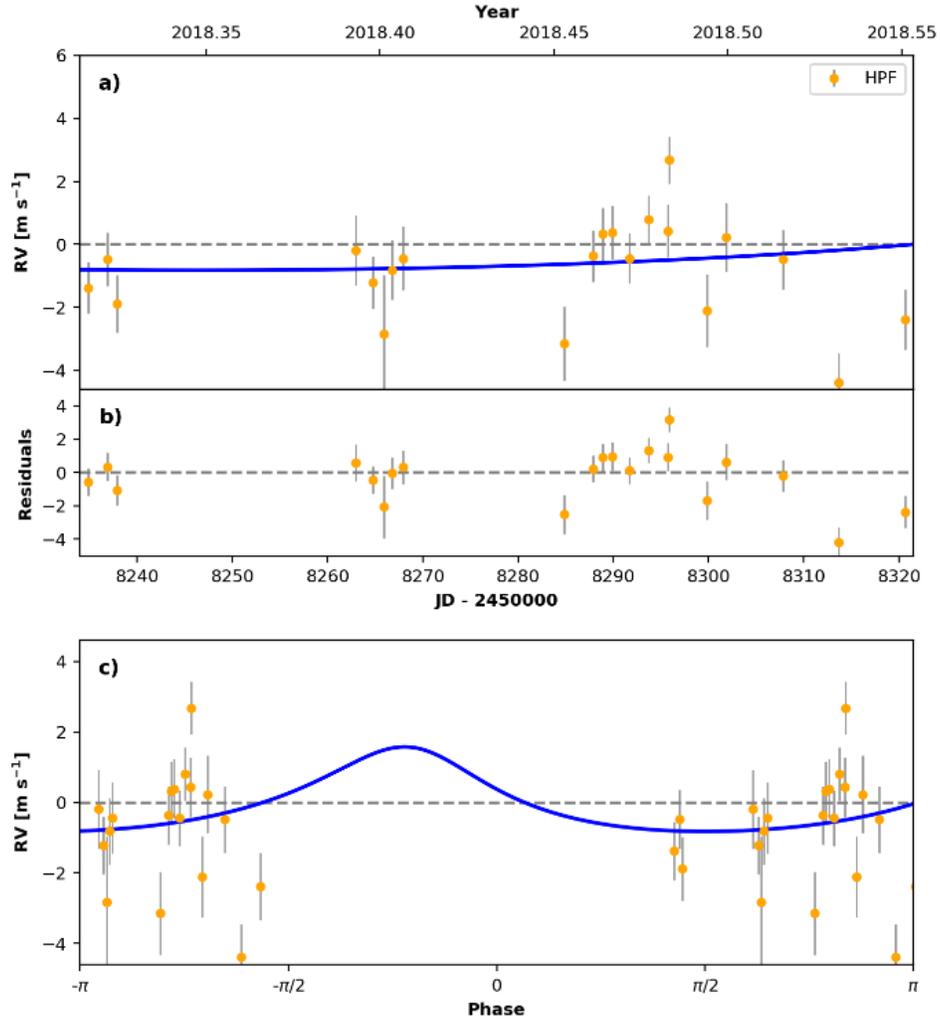

**Figure S2:** a) HPF observations (gold) with the proposed 233-day exoplanet orbit from Ribas et al [21] shown as a solid blue line. We have fit and subtracted a zero-point offset to the HPF velocities. b) Residuals of the HPF RVs after subtracting the exoplanet model. The residuals have an RMS scatter of 1.59 m/s, statistically indistinguishable from our baseline 1.53 m/s stability. c) HPF RVs phased to the proposed 233-day exoplanet period. Figures generated using RadVel [22].

### 4. Consistency of HPF RVs with the Barnard's star exoplanet candidate

Ribas et al. [21] reported the detection of an exoplanet candidate with orbital period 233 days and RV amplitude 1.2 m/s around Barnard's star. While our RV stability should be sufficient to detect such a planet, our 86 day observing baseline coincided with the flattest part of the proposed eccentric orbit for Barnard b, resulting in the stable RV curve shown in Figure 3c. In Figure S2, we show our HPF RVs alongside the Ribas et al. orbital model. We have modeled and subtracted a zero-point offset to the HPF RVs, but have not otherwise modified the RVs or the planet model. When comparing our observations to the phased orbit model, we see that we do not cover the RV maximum, where the candidate planet's eccentric (e = 0.3) orbit creates the greatest deviation from a flat line. Hence, our stability results are essentially identical regardless of the RV model adopted; we find an RMS of 1.53 m/s around the flat-line model, and RMS = 1.59 m/s around the Ribas et al. exoplanet model.

**References**

1. A. J. Metcalf, V. Torres-Company, D. E. Leaird and A. M. Weiner, "High-Power Broadly Tunable Electrooptic Frequency Comb Generator," IEEE Journal of Selected Topics in Quantum Electronics 19, 231-236 (2013).
2. K. Beha, D. C. Cole, P. Del'Haye, A. Coillet, S. A. Diddams, and Scott B. Papp, "Electronic synthesis of light," Optica 4, 406-411 (2017).
3. D. R. Carlson, D. D. Hickstein, W. Zhang, A. J. Metcalf, F. Quinlan, S A. Diddams, S. B. Papp, "Ultrafast electro-optic light with sub-cycle control" to appear in Science (2018).
4. L. C. Sinclair, J.-D. Deschênes, L. Sonderhouse, W. C. Swann, I. Khader, E. Baumann, N.R. Newbury, I. Coddington, "A compact optically coherent fiber frequency comb," Review of Scientific Instruments 86, 081301 (2015).
5. A M. Weiner, "Ultrafast optical pulse shaping: A tutorial review," Optics Communications 294, 3669-3692 (2011).
6. H. Lee, G. J. Hill, B. L. Vattiat, M. P. Smith, M. Haeuser, "Facility calibration unit of Hobby Eberly Telescope wide field upgrade", Proc. SPIE 8444, Ground-based and Airborne Telescopes IV, 84444J (2012).
7. S. Halverson, A. Roy, S. Mahadevan, L. Ramsey, E. Levi, C. Schwab, F. Hearty, N. MacDonald, "An efficient, compact, and versatile fiber double scrambler for high precision radial velocity instruments", The Astrophysical Journal, 806, 61 (2015).
8. M. Shetrone, M. Cornell, J. Fowler, N. Gaffney, B. Laws, J. Mader, C. Mason, S. Odewahn, B. Roman, S. Rostopchin, D. Schneider, J. Umbarger, A. Westfall, " Ten Year Review of Queue Scheduling of the Hobby-Eberly Telescope", PASP, 119, 556 (2007).
9. J. P. Ninan, C. F. Bender, S. Mahadevan, E. B. Ford, A. J. Monson, K. F. Kaplan, R. C. Terrien, A. Roy, P. M. Robertson, S. Kanodia, G. K. Stefansson, "The Habitable-Zone Planet Finder: improved flux image generation algorithms for H2RG up-the-ramp data," Proc. SPIE 10709, High Energy, Optical, and Infrared Detectors for Astronomy VIII, 107092U (2018)
10. K. Horne, "An optimal extraction algorithm for CCD spectroscopy.", Publications of the Astronomical Society of the Pacific 98, 609 (1986).
11. M. Zechmeister, G. Anglada-Escudé, A. Reiners, "Flat-relative optimal extraction. A quick and efficient algorithm for stabilised spectrographs," Astronomy and Astrophysics, 561, A59 (2014).
12. R. C. Terrien, C. F. Bender, S. Mahadevan, S. P. Halverson, L. W. Ramsey, F. R. Hearty, "Developments in simulations and software for a near-infrared precision radial velocity spectrograph," Proc. SPIE 9152, Software and Cyberinfrastructure for Astronomy III, 915226 (2014)
13. G. Anglada-Escudé, R. P. Butler, "The HARPS-TERRA Project. I. Description of the Algorithms, Performance, and New Measurements on a Few Remarkable Stars Observed by HARPS", Astrophysical Journal Supplements, 200, 15 (2012).
14. M. Zechmeister, A. Reiners, P. J. Amado, M. Azzaro,F. F. Bauer,V. J. S. Béjar,J. A. Caballero, E. W. Guenther, H. J. Hagen, S. V. Jeffers, A. Kaminski, M. Kürster, R. Launhardt, D. Montes, J. C. Morales, A. Quirrenbach, S. Reffert, I. Ribas, W. Seifert, L. Tal-Or, V. Wolthoff, "Spectrum radial velocity analyser (SERVAL). High-precision radial velocities and two alternative spectral indicators", Astronomy & Astrophysics, 609, 12 (2018).
15. S. Kanodia, J. Wright, "Python Leap Second Management and Implementation of Precise Barycentric Correction (barycorrpy)", Research Notes of the American Astronomical Society, 2, 4 (2018).
16. K. Gullikson, S. Dodson-Robinson, A. Kraus, "Correcting for Telluric Absorption: Methods, Case Studies, and Release of the TelFit Code", Astronomical Journal, 148, 53 (2014).
17. S. A. Clough, M. W. Shephard, E. J. Mlawer, J. S. Delamere, M. J. Iacono, K. Cady-Pereira, S. Boukabara, P. D. Brown. "Atmospheric radiative transfer modeling: a summary of the AER codes", Journal of Quantitative Spectroscopy and Radiative Transfer, 91, 233-244 (2005).
18. P. Robertson, C. Bender, S. Mahadevan, A. Roy, L. W. Ramsey, "Proxima Centauri as a Benchmark for Stellar Activity Indicators in the Near-infrared", Astrophysical Journal, 832, 112 (2016).
19. M. Kürster, M. Endl, F. Rouesnel, S. Els, S. Briallant, A. P. Hatzes, S. H. Saar, W. D. Cochran, "The low-level radial velocity variability in Barnard's star (= GJ 699). Secular acceleration, indications for convective redshift, and planet mass limits", Astronomy & Astrophysics, 403, 1077 (2003).
20. F. Bouchy, F. Pepe, D.Queloz, "Fundamental photon noise limit to radial velocity measurements", Astronomy & Astrophysics, 374, 733 (2001).
21. I. Ribas, M. Tuomi, A. Reiners, R. P. Butler, J. C. Morales, M. Perger, S. Dreizler, C. Rodríguez-López, J. I. González Hernández, A. Rosich, F. Feng, T. Trifonov, S. S. Vogt, J. A. Caballero, A. Hatzes, E. Herrero, S. V. Jeffers, M. Lafarga, F. Murgas, R. P. Nelson, E. Rodríguez, J. B. P. Strachan, L. Tal-Or, J. Teske, B. Toledo-Padrón, M. Zechmeister, A.

Quirrenbach, P. J. Amado, M. Azzaro, V. J. S. Béjar, J. R. Barnes, Z. M. Berdiñas, J. Burt, G. Coleman, M. Cortés-Contreras, J. Crane, S. G. Engle, E. F. Guinan, C. A. Haswell, Th. Henning, B. Holden, J. Jenkins, H. R. A. Jones, A. Kaminski, M. Kiraga, M. Kürster, M. H. Lee, M. J. López-González, D. Montes, J. Morin, A. Ofir, E. Pallé, R. Rebolo, S. Reffert, A. Schweitzer, W. Seifert, S. A. Shectman, D. Staab, R. A. Street, A. Suárez Mascareño, Y. Tsapras, S. X. Wang & G. Anglada-Escudé, "A candidate super-Earth planet orbiting near the snow line of Barnard's star," Nature 563, 365–368 (2018).
22. B. J. Fulton, E. A. Petigura, S. Blunt, E Sinukoff, "RadVel: The Radial Velocity Modeling Toolkit," Publications of the Astronomical Society of the Pacific, Volume 130, Issue 986, pp. 044504 (2018)